\documentclass[aps,prl,twocolumn,showpacs,superscriptaddress,nofootinbib,preprintnumbers]{revtex4-1}

\usepackage{style}
\usepackage{colortbl}
\usepackage{multirow}
\usepackage{graphicx}
\usepackage{amsmath}
\usepackage{amssymb}
\usepackage{amsthm}
\usepackage{amsfonts}
\usepackage{braket}
\usepackage{slashed}
\usepackage[T1]{fontenc}
\usepackage{mathtools}
 
\usepackage{tikz}
\usepackage{tikz-feynman}
\usetikzlibrary{decorations.pathreplacing}

\usepackage{ytableau}

\definecolor{newgray}{gray}{0.95}

\begin{document}

\preprint{KEK-QUP-2026-0008, KEK-TH-2836}

\title{Echoes of Nucleon Decay from Long-Lived Particles 
} 

\author{Patrick Adolf} 
\email{patrick.adolf@tu-dortmund.de}
\affiliation{Fakult\"at f\"ur Physik, Technische Universit\"at Dortmund,
  D-44221 Dortmund, Germany}

\author{Chandan Hati} 
\email{chandan@ific.uv.es}
\affiliation{Instituto de Física Corpuscular (IFIC), Universitat
  de València-CSIC,
C/ Catedratico Jose Beltran, 2, E-46980 Valencia, Spain}

\author{Martin Hirsch} 
\email{mahirsch@ific.uv.es}
\affiliation{Instituto de Física Corpuscular (IFIC), Universitat
  de València-CSIC,
C/ Catedratico Jose Beltran, 2, E-46980 Valencia, Spain}

\author{Volodymyr Takhistov}
\email{vtakhist@post.kek.jp}
\affiliation{International Center for Quantum-field Measurement Systems
  for Studies of the Universe
and Particles (QUP), KEK, 1-1 Oho, Tsukuba, Ibaraki 305-0801, Japan
}
\affiliation{Theory Center, Institute of Particle and Nuclear Studies,
  High Energy Accelerator Research Organization (KEK), Tsukuba 305-0801, Japan}
\affiliation{Graduate University for Advanced Studies (SOKENDAI),
1-1 Oho, Tsukuba, Ibaraki 305-0801, Japan
}
\affiliation{Kavli Institute for the Physics and Mathematics of the Universe
  (WPI), Chiba 277-8583, Japan}

\begin{abstract} 
Nucleon decay searches provide uniquely sensitive probes of baryon number violation and physics beyond the Standard Model. We propose a new class of nucleon decay observables involving long-lived particles (LLPs), characterized by spatially separated but temporally correlated ``echo'' vertices not captured by conventional prompt searches. Focusing on vector LLPs, we construct effective operators and ultraviolet realizations, and show that Super-Kamiokande, Hyper-Kamiokande and JUNO can achieve geometric acceptances approaching 80\% over a broad range of LLP decay lengths.  Echo signatures could in principle
arise from any visibly decaying LLP.
 \end{abstract}

\maketitle

\raggedbottom

\paragraph{\textbf{Introduction.}\label{sect:intro} }

Baryon number violation is a generic prediction of  many  scenarios beyond the Standard
Model (SM), where the baryon number is an accidental symmetry. It is also a
necessary ingredient for explaining the observed matter-antimatter asymmetry of
the universe~\cite{Sakharov:1967dj}. The extraordinary stability of 
matter suggests that such violations should occur at energy scales far beyond
those directly accessible in laboratories, including colliders. Nucleon decays
thus provide uniquely sensitive probes of fundamental physics at extreme
energies. Here we show that nucleon decays involving long-lived particles (LLPs)
give rise to qualitatively new class of observable signatures, providing powerful probes of baryon number
violation and new physics.

While nucleon decay has been extensively explored in contexts such as
Grand Unified Theories (GUTs)~\cite{Georgi:1974sy,Fritzsch:1974nn},
experimental searches have largely focused on prompt final states
with SM particles, such as $p\rightarrow
e^+\pi^0$~\cite{Super-Kamiokande:2020wjk} (see~\cite{Takhistov:2026aln} for a
review), with only few dedicated searches performed for non-standard
topologies involving missing energy final
states~\cite{Super-Kamiokande:2015pys}. More broadly, a wide range of
weakly coupled states, such as sterile neutrinos and dark
fermions~\cite{Davoudiasl:2014gfa,Fornal:2018eol,McKeen:2018xwc,Helo:2018bgb,Heeck:2025uwh},
can arise in motivated SM extensions and lead to non-canonical nucleon
decay topologies. Building on the general framework developed in
Ref.~\cite{Fridell:2023tpb}, such signatures can be misidentified or
entirely escape conventional searches.  

\begin{figure}[t]
    \centering \includegraphics[scale=0.9]{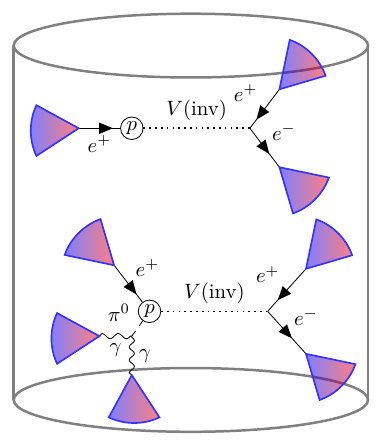}
    \caption{Illustration
    of nucleon decays with ``echo'' signature in a detector for two
    representative processes $p \to e^+V$ (Top) and $p \to e^+ \pi^0
    V$ (Bottom). The vector LLP, $V$, decays to a lepton pair $e^+e^-$
    after traversing a macroscopic distance and with a time delay
    relative to the parent nucleon decay.  \label{fig:DblBng}}
\end{figure}

In this Letter, we propose a qualitatively new class of nucleon decay
observables involving LLPs, which naturally arise in many SM
extensions~\cite{Alimena:2019zri,Agrawal:2021dbo}. A particularly well
motivated class are light vector mediators such as dark photons and
$Z'$ bosons~\cite{Ilten:2018crw}.  Existing experimental
constraints~\cite{Batell:2022dpx} on the couplings of light
vectors to SM particles motivate weakly coupled scenarios in which
macroscopic decay lengths naturally arise for
$m_V \lesssim \mathcal{O}(1)\mathrm{GeV}$.  While LLPs produced in
nucleon decay have previously been considered in specific
model-dependent scenarios of R-parity violating
supersymmetry~\cite{Domingo:2024qoj}, here we develop a general framework for
displaced ``echo'' signatures. These novel topologies feature
spatially separated but temporally correlated vertices, arising when
LLPs propagate macroscopic distances before decaying visibly inside
the detector, as illustrated in Fig.~\ref{fig:DblBng}.

While delayed nucleon decay signatures have been explored before with
SM final states, such as in $p \to \nu K^+$ searches where an
invisible kaon decays after a delay determined by its SM
lifetime~\cite{Super-Kamiokande:2014otb}, the echo topology we put
forth is qualitatively distinct. Here, the LLP decay products are
visible and produce identifiable displaced vertices, with a continuous
distribution of spatial separations governed by fundamental couplings
rather than by the lifetime of a known SM particle. Such signatures
are not captured by conventional prompt nucleon decay searches and
open a new avenue for probing baryon number violation and new physics
in existing experiments such as Super-Kamiokande
(Super-K)~\cite{Abe:2013gga} and JUNO~\cite{JUNO:2021vlw}, as well as
upcoming detectors such as Hyper-Kamiokande
(Hyper-K)~\cite{Hyper-Kamiokande:2018ofw} and
DUNE~\cite{DUNE:2020lwj}.

\paragraph{\textbf{Long-lived particles from nucleon decays.}\label{sect:eft}}  
 
Weakly coupled light novel fields beyond the SM can
naturally give rise to LLP final states in nucleon decays. Motivated examples include axion-like
particles, sterile neutrinos and vector bosons such
as dark photons or $Z'$ gauge bosons associated with an
additional $\text{U}(1)'$ symmetry.

For concreteness, we focus on light vector bosons, which arise naturally in many SM 
extensions and ultraviolet completions. Nucleon decay kinematics require the vector $V^{\mu}$ mass to be
$m_V \lesssim \mathcal{O}(1)$~GeV. Existing constraints including from laboratory, astrophysics and precision measurements typically restrict respective vector couplings to SM constituents  to be 
$g \lesssim 10^{-3}$~\cite{Batell:2022dpx}. This naturally enables macroscopic 
decay lengths over a significant region of parameter space.

Such vectors can be minimally realized, for example, in scenarios where a broken
$\text{U}(1)'$ gauge symmetry yields a $Z'$ with mass generated 
by a singlet scalar field vacuum expectation value, with SM interactions arising 
through direct gauge couplings and kinetic mixing with hypercharge. The
physical state couples to SM currents with effective strength, up to model dependent factors. Yet another minimal realization that provides closely resembling features is a dark 
photon coupled solely via kinetic mixing~\cite{Holdom:1985ag,Ilten:2018crw,Agrawal:2021dbo}.
In both cases, existing constraints permit parameter regions yielding observable 
echo signatures complementary to conventional searches. More so, such light vector states can 
also emerge in GUT frameworks when an additional $\text{U}(1)'$ survives as an 
intermediate symmetry below the unification scale, with small couplings arising 
from gauge coupling suppression, kinetic mixing or hidden sector sequestering.
In the Supplemental Material, 
we provide an explicit GUT realization of such scenario.

More generally, echo signatures can be analyzed within the model independent framework 
of SM gauge group-invariant effective field theory (SMEFT), extended to include additional light vector states that appear below the energy scale $\Lambda$. 
In the Supplemental Material, we classify the leading dimension
$d=7$ and $d=8$ operators,  $\mathcal{O}$, containing a single covariant
derivative and violating baryon (and lepton) number.
As a representative minimal example, consider the two-body proton decay
 $p\rightarrow e^+V$ with $e^+$ being the positron, as displayed in Fig.~\ref{fig:DblBng},
generated at $d=8$ by 
\begin{equation}{\label{eq:EFO}}
 \mathcal{L}\supset\frac{C_{Heu^2Q}}{\Lambda^4}
       \mathcal{O}_{Heu^2Q}\equiv
 \frac{C_{Heu^2Q}}{\Lambda^4}H^{\dagger i} (\overline{e^c}
 u)(\overline{u^c}\gamma_\mu Q_{i}) V^\mu\,,
\end{equation}
where $H$ denotes the SM Higgs doublet, $Q_i$ the
left handed quark doublet with $\text{SU}(2)_L$ index $i$,
$u$ and $e$ the right handed up quark and electron
fields. Here, $C_{Heu^2Q}$ is a dimensionless Wilson
coefficient and $\Lambda$ denotes the heavy new physics
scale. Generation indices for SM fermions have been
suppressed.

In conventional nucleon decays involving pseudoscalar
mesons and leptons, the relevant amplitudes can be expressed by proton decay matrix elements
with a Lorentz scalar structure involving
three quarks of different chiralities~\cite{Claudson:1981gh}, which
have been calculated by ab initio lattice QCD methods~\cite{JLQCD:1999dld,Aoki:2017puj}.  In contrast,
decays involving a vector final state as described here are proportional to the
polarization  $\epsilon^{\ast\mu}$ of the outgoing vector
$V^\mu$ and therefore require an overall Lorentz vector
structure from the remaining amplitude component. Consider two-body decay
$p(p)\rightarrow e^+(k)V(q)$, where the arguments in the parenthesis denote the different momenta. Momentum conservation and
transversality of the vector polarization imply that the
general amplitude can be expressed in terms of independent Lorentz structures as
\begin{equation}{\label{eq:FF2b}} {\mathcal{M}}= \bar{u}_{e}
 P_{R}\left[ F_1(q^2) \gamma^{\mu}  + F_2(q^2 )  i \sigma^{\mu\nu} q_\nu
 \right] u_{p} \epsilon^{\ast}_{\mu}\,,
\end{equation}
where $u_p$ and $u_e$ denote the proton and positron
spinors, respectively, $P_R$ is the
right handed chiral projection operator, $\sigma^{\mu\nu}$ is the antisymmetric tensor matrix, $q=p-k$ is the outgoing vector
momentum, and $F_1$, $F_2$ are form factors encoding the
hadronic transition matrix elements.

At present, no ab initio lattice calculation exists for
such vector-current matrix elements. Instead, we employ recently developed methods~\cite{Liao:2025vlj,Song:2026gyo} for estimating the relevant low energy constants based on dimensional
analysis~\cite{Weinberg:1989dx,Manohar:1983md} in
leading order chiral perturbation theory.
 
To estimate the relevant form factors entering
Eq.~\eqref{eq:FF2b}, we perform the chiral matching. The form factors in
Eq.~\eqref{eq:FF2b} can be parametrized as
$F_1=C(\Lambda,\Lambda_\chi)m_p/4$ and
$F_2=C(\Lambda,\Lambda_\chi)/2$, where
$\Lambda_\chi\sim1.2~{\rm GeV}$ denotes the chiral
symmetry breaking scale, $m_p$ is the proton mass, and
light SM lepton masses have been neglected. Matching 
the effective operator in Eq.~\eqref{eq:EFO} to chiral operators, we obtain
$ C(\Lambda,\Lambda_\chi)=
c(\Lambda_\chi) C_{Heu^2Q} v/(
\Lambda_\chi \Lambda^4 \sqrt{2})$,
 where the low energy constant is given as
$c(\Lambda_\chi)\sim \Lambda_\chi^3/(4\pi)^2 \simeq
0.011~\text{GeV}^3$~\cite{Liao:2025vlj}. Here, $v$ is the Higgs vacuum expectation value.

Using the resulting decay amplitude and standard techniques, we can calculate the benchmark
partial lifetimes for the channels
enabling sensitivity to echo signatures. For the example 
two-body decay mode
$p\rightarrow e^+V$ the resulting half-life can be expressed in the simple form 
\begin{equation}\label{eq:t12_2b}
T_{1/2} \simeq 1.6 \times 10^{35}
\left(\frac{\Lambda}{10^8 \ {\rm GeV}}\right)^8 f(x)\ \text{yr}~,
\end{equation}
where we have neglected the light SM lepton mass, and the function
$f(x) = {x}/{[(1-x)^2 ( 1 - 2 x + 4 x^2)]}$ with
$x=(m_V/m_p)^2$. Thus, a half-life of the order $10^{35}$ yr enables sensitivity to heavy new physics scales of the order of $10^{8}$~GeV. While Eq.~\eqref{eq:t12_2b} is derived for the operator
in Eq.~\eqref{eq:EFO}, analogous conclusions hold for
other operators yielding the same final states,
up to numerical factors.

The computation for the three-body decay mode $p \rightarrow \pi^0   e^+ 
V $ displayed in
Fig.~\ref{fig:DblBng}     can be performed analogously, although
the larger number of independent external momenta allows
additional Lorentz structures in the amplitude. In the Supplemental Material, we discuss
the calculation details,
including the chiral matching used to estimate the
relevant low energy constants.

\paragraph{\textbf{Echo signatures and sensitivity reach.}
  \label{sect:exp}}

\begin{figure*}[t] \includegraphics{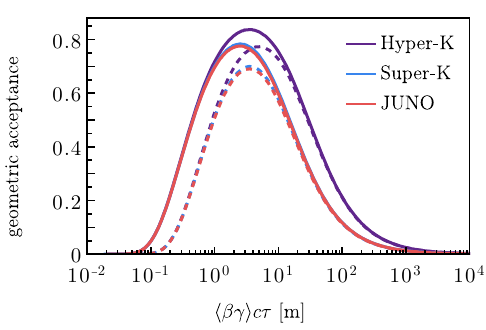}
  \includegraphics{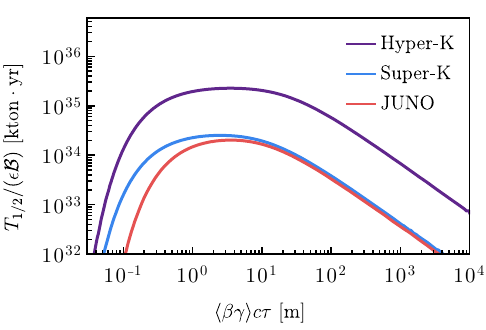}
\caption{Geometric acceptance (Left) and partial lifetime sensitivity 
$T_{1/2}/(\mathcal{B}\epsilon)$ (Right) as a function of $\langle\beta\gamma\rangle c\tau$ 
for Super-K, Hyper-K and JUNO. Two-body decay $p \to e^+ V$ is assumed. 
Left panel results are shown for vertex resolutions $l_{\rm min} = 0.3$~m (solid) 
and $l_{\rm min} = 0.6$~m (dashed) for all detectors. 
Right panel sensitivity reach assumes $l_{\rm min} = 0.3$~m for Super-K and Hyper-K 
and $l_{\rm min} = 0.6$~m for JUNO, with exposures of 225 kton$\cdot$yr for Super-K, 1870 kton$\cdot$yr for Hyper-K, and 200~kton$\cdot$yr for JUNO, corresponding to 10 years of detector livetime.\label{fig:signatures}}
\end{figure*}

The LLP produced in nucleon decay
propagates over a characteristic distance $l = \beta \gamma  c \tau$, where
$\beta$ and $\gamma$ are relativistic factors and $\tau$ is the proper lifetime, 
with a corresponding decay time delay $t_{\rm delay} = l/(\beta c)$  relative to the primary vertex.
Detection of the echo topology requires the secondary LLP decay vertex to lie within the fiducial volume and satisfy $l > l_{\min}$, where $l_{\min}$ is the minimum resolvable vertex separation. Together with the detector size setting $l_{\max}$, this determines the accessible range $  l_{\min} \lesssim l \lesssim l_{\max}$  and hence the region of parameter space probed. Detector
energy thresholds and resolution can affect the detection efficiency of
the visible final states.
 
 We focus primarily on water Cherenkov detectors, which provide leading
sensitivity to a broad range of nucleon decay channels, and consider
scintillator detectors to illustrate complementarity. For the detector geometry
we model Super-K~\cite{Abe:2013gga} as a cylindrical detector with
diameter  $d=39.3 \text{ m}$ and height $h=41.4 \text{ m}$, and apply a fiducial
volume cut of $2 \text{ m}$ from the walls, corresponding to an effective mass
of $22.5 \text{ kton}$ as employed in conventional analyses, although larger
fiducial volumes can be considered with optimized
reconstruction~\cite{Super-Kamiokande:2020wjk}. For Hyper-K, we use
$d=74 \text{ m}$ and $h=60 \text{ m}$, with a fiducial cut of $1.5 \text{ m}$,
yielding an effective mass of approximately $187 \text{
kton}$~\cite{Hyper-Kamiokande:2018ofw}. JUNO is modeled as a spherical detector
with diameter $d=35.4 \text{ m}$ filled with organic liquid scintillator,
corresponding to an effective fiducial mass of about $20 \text{
kton}$~\cite{JUNO:2021vlw}.

For Super-K we adopt benchmark values of $l_{\rm min} \simeq 0.3~{\rm
m} $ for the level of vertex separations that can be resolved in event
reconstruction, and timing resolutions of order $ t \sim 1~{\rm
ns}$~\cite{Abe:2013gga}. Comparable performance is expected to be achieved at
Hyper-K~\cite{Hyper-Kamiokande:2018ofw}, with similar detection
principles.  For liquid scintillator detectors such as JUNO, the reconstruction
characteristics differ due to photon propagation and the extended nature of
electromagnetic showers. The timing resolution depends on the photodetector
system and is expected to be at the level of a few nanoseconds. We adopt $ t
\sim 3$~ns as benchmark~\cite{JUNO:2021vlw}. At higher energies that
are typically relevant for nucleon decays the effective spatial resolution is
expected to be broadened and we therefore adopt a conservative estimate of
$l_{\rm min} \simeq 0.6\ {\rm m}$ for JUNO~\cite{JUNO:2021vlw}.

We perform simulations of nucleon decay events with subsequent LLP decays for
each detector configuration described above to characterize the expected distribution of displaced ``echo'' vertices. Primary
nucleon decays are generated assuming a uniform spatial distribution within the
fiducial volume.  The LLP is then propagated according to its boost factor and
proper lifetime  with decay length  sampled from the corresponding exponential
distribution in the laboratory frame. The simulations are performed at the
event truth level and do not include detector response or reconstruction
effects. In particular, we do not model the detailed topology of LLP decay
products, but instead focus on the spatial and temporal separation of the
primary and secondary vertices.  This allows us to estimate the geometric acceptance under detector-specific constraints on vertex and timing resolution.

We define the geometric acceptance factor $a$ as the probability that a nucleon
decay occurring within the fiducial volume is followed by an LLP decay that also
occurs within the fiducial volume and satisfies $l > l_{\rm min}$. We emphasize
that this quantity is distinct from the overall signal detection efficiency
$\epsilon$. Fig.~\ref{fig:signatures} displays geometric acceptance as a function of $\langle
\beta\gamma\rangle c\tau$ for the different experiments. The results shown
assume two-body decays of free protons with fixed $\langle \beta\gamma\rangle$.
For bound nucleons, Fermi motion induces a smearing of $\beta\gamma$, and
three-body decays lead to broader kinematic distributions. We have estimated (see Supplemental Material for details) 
that these effects modify the acceptance at the level of $\lesssim 2\%$ and
$\lesssim 10\%$, respectively, over the range $\langle \beta\gamma\rangle c\tau
= [0.1,\,500]~{\rm m}$, and therefore do not qualitatively affect our
conclusions. The maximal acceptance
is approximately $77\%$ ($84\%$) for
Super-K (Hyper-K).

Estimating the reach of the different experiments requires, besides geometric acceptance, the overall detection efficiency $\epsilon$. The latter depends
on detector response and reconstruction, which for reliable determination
requires dedicated detector level simulations, beyond the scope of this work.
Our results therefore provide approximate benchmark geometric estimates of the
accessible parameter space, to be convolved with efficiencies.
We note that Super-K has searched for prompt proton decay into three charged leptons, reporting  $\epsilon \simeq 0.6$   for $  p \to e^+e^+e^-$~\cite{Super-Kamiokande:2020tor}. While not directly applicable to the displaced echo topology here, this search is conceptually related in the limit $  l \lesssim l_{\rm min}$, where the echo signal would be  reconstructed as a prompt trilepton final state.

We parametrize the partial nucleon decay half-life reach for the echo signature
as  
\begin{equation}
    \dfrac{T_{1/2} }{ \epsilon \mathcal{B}} \simeq \dfrac{\lambda N_\text{T}   a(\beta \gamma c\tau)}{N_{\rm CL}}\,,
\end{equation}
where $\mathcal{B}$ is the branching ratio, $N_\text{T}$ is the number
of relevant target nucleons per kiloton, $\lambda$ is the experimental
exposure in kton-years, $\epsilon$ is the signal efficiency, and $N_{\rm CL}$ is the 90\% C.L.
upper limit on the number of signal events, where we take $N_{\rm CL} = 2.3$ for a background free estimate.

In Fig.~\ref{fig:signatures} we display the corresponding sensitivity reach as a function of $\langle \beta\gamma\rangle c\tau$, factoring out signal efficiencies and assuming that echo signatures are clearly distinguishable and essentially free from backgrounds. The structure of the curves
follows directly from the geometric acceptance in Fig.~\ref{fig:signatures}. For all detectors we consider all available target protons, including protons bound in nuclei. The
reach is suppressed for decay lengths shorter than the minimum resolvable vertex
separation, rising when the secondary LLP decay is typically contained and
displaced within the fiducial volume  and decreasing again once the LLP decay
length becomes comparable to or larger than the detector size scale. The results
should therefore be interpreted as  geometry-level sensitivity estimates.
Determining the efficiency for echo events requires dedicated reconstruction and background studies by experimental collaborations.

 Importantly, echo signatures probe parameter space complementary to conventional searches for light vector mediators, as illustrated for dark photons (see Supplemental Material for details). Unlike typical dark photon experiments, where both production and decay are controlled by the electromagnetic field strength kinetic mixing parameter $\epsilon$, nucleon decay echo signals involve a baryon number-violating coupling at production while $\epsilon$ governs only the LLP decay. This distinction enables nucleon decay experiments to probe regions of the $(\epsilon,m_V)$ parameter space corresponding to very small $\epsilon$ and hence long decay lengths, beyond the projected reach of proposed dark photon searches and complementary to proposed experiments such as SHIP~\cite{Zhou:2024aeu}, FASER-2~\cite{FASER:2018eoc}, DarkQuest~\cite{Berlin:2018pwi} and DUNE near detector~\cite{Berryman:2019dme}. Conversely, for larger kinetic mixing, potential echo signals at experiments such as Hyper-K could be cross-validated by future dark photon searches.

\paragraph{\textbf{Conclusions.}\label{sect:cncl} } 

We have identified a qualitatively new class of nucleon decay
signatures involving LLPs, characterized by spatially separated but
temporally correlated displaced ``echo'' vertices that are generally
not captured by conventional prompt searches. Focusing on vector LLPs,
we presented effective operators and possible ultraviolet realizations giving
rise to such novel signals, and demonstrated that Super-K, Hyper-K and JUNO
experiments can achieve geometric acceptances approaching 80\% for
representative decay channels over broad LLP decay lengths. Echo
signatures additionally probe regions of light vector parameter space
complementary to conventional dark photon searches, including small
kinetic mixing corresponding to long decay lengths. While we focused
on vector LLPs, displaced echo signatures can in principle arise from
any visibly decaying LLP, including scenarios with multiple LLPs
leading to more complex echo topologies.  This broadly extends
nucleon decay searches beyond standard channels and opens a new avenue
for probing baryon number violation and light new physics in current
and future experiments.

\begin{acknowledgments}
\paragraph{\textbf{Acknowledgments.}---}
P.~A. is supported by the \textit{Studienstiftung des deutschen Volkes} and
thanks the IFIC for hospitality during the initial phase of this project.  C.~H. is funded by the Generalitat Valenciana under Plan Gen-T via CDEIGENT grant No. CIDEIG/2022/16. This work is support by the Spanish grants PID2023-147306NB-I00 and CEX2023-001292-S (MCIU/AEI/10.13039/501100011033). This work is also supported in part by the World Premier International Research
Center Initiative (WPI), MEXT, Japan.
\end{acknowledgments}

\setlength{\bibsep}{3pt} 
\bibliography{biblio}

\clearpage
\newpage  
\appendix
\onecolumngrid
\renewcommand{\theequation}{S.\arabic{equation}}
\setcounter{equation}{0}

\centerline{\large {Supplemental Material for}}
\medskip

{\centerline{\large \bf{Echoes of Nucleon Decay from Long-Lived Particles}}}
\medskip
{\centerline{Patrick Adolf, Chandan Hati, Martin Hirsch, Volodymyr Takhistov}}
\bigskip
\bigskip
  
\vspace{1em}  

\begin{appendix}   
\onecolumngrid 

In this Supplemental Material we provide additional details on the EFT operators with LLPs, UV completions, nucleon decay kinematics and geometric acceptance computation.

 \subsection{A. Effective Field Theory Operators\label{app:EFT}}

\begin{table}[h]
\renewcommand*{\arraystretch}{1.2}
\centering
\ytableausetup
{boxsize=1.em}
\begin{tabular}{|c|c|c|c|c|c|}
  \hline
Notation & Operator & $n_G=3$ & Symmetry & Proton Dec. & Neutron Dec. \\
  \hline
      ${\cal O}_{ed^3V}$ &  
      $(\overline{e_{R,p}}\gamma^{\mu}d_{R,r})(\overline{d_{R,s}^c}D_{\mu}d_{R,t})$ &
      30 &
\ytableausetup{centertableaux}
\begin{ytableau}
       \none & r & s & t & \none \\
\end{ytableau}
  &   $-$ & $\pi^+ e^- V$ \\
\hline
      ${\cal O}_{LQd^2V}$ &  
      $(\overline{L_p}\gamma^{\mu}Q_r)(\overline{d_{R,s}^c}D_{\mu}d_{R,t})$ &
      54 &
\ytableausetup{centertableaux}
\begin{ytableau}
       \none & s &t  & \none 
\end{ytableau}
  &    $\pi^+\nu V$ & $\nu V$, $\pi^+e^- V$, $\pi^0\nu V$  \\
      \hline
\end{tabular}
\caption{ SMEFT operators containing a single gauge vector at $d=7$
  that violate baryon and lepton number and can induce nucleon
  decay. Operator short-hand notation and corresponding operator
  structure in four-component notation are shown, with generation
  indices written explicitly. The column $n_G=3$ gives the number of
  independent operator coefficients assuming three generations of
  fermions. The symmetry column denotes properties in the flavour
  indices of repeated fermion fields encoded with Young tableaux.
  Representative proton and neutron decay channels induced by the
  corresponding operators are displayed. The vector field $V$ is
  contained in the covariant derivative $D_\mu$. The operator ${\cal
    O}_{ed^3V}$ is a special case, see the discussion in the text.}
\label{tab:Opd7}
\end{table}

Here, we discuss EFT operators at
$d=7$ and $d=8$ relevant for nucleon decay involving novel vector
states $V$. For the construction of the operator basis we use computer code packages~\texttt{AutoEFT}~\cite{Harlander:2023ozs} and
\texttt{Sym2Int}~\cite{Fonseca:2017lem}. While the operator tables
presented below largely follow the conventions of
Ref.~\cite{Harlander:2023ozs}, we convert the output of
\texttt{AutoEFT} to four-component notation. Using 
\texttt{Sym2Int} we cross-checked the parameter counting and symmetry
properties of the operators and verified that there is a complete agreement between the
two codes for all operators considered.

We focus exclusively on operators that violate baryon (and lepton)
number and can therefore induce nucleon decays. From the complete set
of possible operators, for simplicity, we further select those containing exactly one exotic
vector state. Throughout, we assume that the vector originates from a
spontaneously broken local gauge symmetry. Note that the software code
\texttt{AutoEFT}~\cite{Harlander:2023ozs} does not allow explicit
addition of vector fields to the particle content. Instead, additional
gauge vectors are incorporated through extensions of the SM gauge
group. In contrast, \texttt{Sym2Int}~\cite{Fonseca:2017lem} 
allows vector fields to be introduced directly without requiring a
gauge interpretation. We briefly comment on the differences between
these two constructions at the end of this section.

In Tab.~\ref{tab:Opd7} we present the list of baryon number-violating
operators at $d=7$ in the context of our extended SMEFT, including a
light vector, together with the number of independent operator
coefficients for three generations of fermions and the corresponding
nucleon decay final states, both protons and neutrons. The exotic
gauge vector is contained in the covariant derivative $D_{\mu}$. The
$\mathrm{SU}(3)$ contractions are always of the form
$\epsilon_{\alpha\beta\gamma} q_1^\alpha q_2^\beta q_3^\gamma$ and we
therefore do not display them explicitly. For the operators shown in
Tab.~\ref{tab:Opd7}, there is only a single possible $\mathrm{SU}(2)$
contraction, which we likewise suppress. For simplicity, we explicitly
display only electron $e$ as a lepton final states, although muon
final states are also allowed whenever kinematically accessible.

In Tab.~\ref{tab:Opd8} we list the relevant operators at $d=8$. All
operators shown are of the type $\psi^4 D H$. At this order there also
exist operators of the form $\psi^4 D^2$. However, such operators are
phenomenologically less interesting in the present context since
they are necessarily accompanied by $\psi^4$ operators inducing proton
decay already at $d=6$. We therefore do not list the $\psi^4 D^2$
operators explicitly.

\begin{table}[!hbt]
\renewcommand*{\arraystretch}{1.0}
\centering
\ytableausetup
{boxsize=0.8em}
\begin{tabular}{|c|c|c|c|c|c|}
  \hline
Notation & Operator & $n_G=3$ & Symmetry & Proton Dec. & Neutron Dec. \\
  \hline
      ${\cal O}_{HLdu^2V}$ &  & 135 & & 
      $e^+V$, $\pi^0e^+V$ & $\pi^-e^+ V$ \\
  1 &  $\epsilon_{ij}H^\dagger_i(\overline{L_{p}^{j,c}}\gamma^{\mu}u_{R,s})
        (\overline{u_{R,t}^c}D_{\mu}d_{R,r})$ &
      54 &
 \ytableausetup{nocentertableaux}
\begin{ytableau}
       \none & s & t & \none \\
 \end{ytableau}
&  & \\
  2 &  $\epsilon_{ij}H^\dagger_i(\overline{L_{p}^{j,c}}\gamma^{\mu}d_{R,r})
  \{(\overline{u_{R,s}^c}D_{\mu}u_{R,t}) + (\overline{u_{R,t}^c}D_{\mu}u_{R,s})\}$ &
      54 &
 \ytableausetup{nocentertableaux}
\begin{ytableau}
       \none & s & t & \none \\
 \end{ytableau}
&  & \\
  3 &  $\epsilon_{ij}H^\dagger_i(\overline{L_{p}^{j,c}}\gamma^{\mu}d_{R,r})
  \{(\overline{u_{R,s}^c}D_{\mu}u_{R,t}) - (\overline{u_{R,t}^c}D_{\mu}u_{R,s})\}$ &
      27 &
 \ytableausetup{centertableaux}
 \begin{ytableau}
       \none & s & \none \\
       \none & t & \none \\
\end{ytableau} 
&  & \\
  \hline
      ${\cal O}_{HeQu^2V}$ &  & 135 & & 
      $e^+V$, $\pi^0e^+V$ & $\pi^-e^+ V$ \\
  1 &  $\epsilon_{ij}H^\dagger_i(\overline{Q_{r}^{j,c}}\gamma^{\mu}u_{R,s})
        (\overline{u_{R,t}^c}D_{\mu}e_{R,p})$ &
      54 &
 \ytableausetup{nocentertableaux}
\begin{ytableau}
       \none & s & t & \none \\
 \end{ytableau}
&  & \\
  2 &  $\epsilon_{ij}H^\dagger_i(\overline{Q_{r}^{j,c}}\gamma^{\mu}e_{R,p})
  \{(\overline{u_{R,s}^c}D_{\mu}u_{R,t}) + (\overline{u_{R,t}^c}D_{\mu}u_{R,s})\}$ &
      54 &
 \ytableausetup{nocentertableaux}
\begin{ytableau}
       \none & s & t & \none \\
 \end{ytableau}
&  & \\
  3 &  $\epsilon_{ij}H^\dagger_i(\overline{Q_{r}^{j,c}}\gamma^{\mu}e_{R,p})
  \{(\overline{u_{R,s}^c}D_{\mu}u_{R,t}) - (\overline{u_{R,t}^c}D_{\mu}u_{R,s})\}$ &
      27 &
 \ytableausetup{centertableaux}
 \begin{ytableau}
       \none & s & \none \\
       \none & t & \none \\
\end{ytableau}
&  & \\
\hline
      ${\cal O}_{HLud^2V}$ &  & 135 & & 
      $\pi^+\overline{\nu}V$ & $\overline{\nu} V$, $\pi^0\overline{\nu} V$ \\
  1 &  $\epsilon_{ij}H^i(\overline{L_{p}^{j,c}}\gamma^{\mu}d_{R,s})
        (\overline{d_{R,t}^c}D_{\mu}u_{R,r})$ &
      54 &
 \ytableausetup{nocentertableaux}
\begin{ytableau}
       \none & s & t & \none \\
 \end{ytableau}
&  & \\
  2 &  $\epsilon_{ij}H^i(\overline{L_{p}^{j,c}}\gamma^{\mu}u_{R,r})
  \{(\overline{d_{R,s}^c}D_{\mu}d_{R,t}) + (\overline{d_{R,t}^c}D_{\mu}d_{R,s})\}$ &
      54 &
 \ytableausetup{nocentertableaux}
\begin{ytableau}
       \none & s & t & \none \\
 \end{ytableau}
&  & \\
  3 &  $\epsilon_{ij}H^i(\overline{L_{p}^{j,c}}\gamma^{\mu}u_{R,r})
  \{(\overline{d_{R,s}^c}D_{\mu}d_{R,t}) - (\overline{d_{R,t}^c}D_{\mu}d_{R,s})\}$ &
      27 &
 \ytableausetup{centertableaux}
 \begin{ytableau}
       \none & s & \none \\
       \none & t & \none \\
\end{ytableau}
&  & \\
\hline
      ${\cal O}_{HeQduV}$ &  & 243 &   & 
      $e^+V$, $\pi^0e^+V$ & $\pi^-e^+ V$ \\
  1 &  $\epsilon_{ij}H^i(\overline{Q_{r}^{j,c}}\gamma^{\mu}e_{R,p})
        (\overline{u_{R,s}^c}D_{\mu}d_{R,t})$ &
      81 & & & \\
  2 &  $\epsilon_{ij}H^i(\overline{Q_{r}^{j,c}}\gamma^{\mu}d_{R,t})
        (\overline{u_{R,s}^c}D_{\mu}e_{R,p})$ &
      81 & & & \\
  3 &  $\epsilon_{ij}H^i(\overline{Q_{r}^{j,c}}\gamma^{\mu}d_{R,t})
        (\overline{e_{R,p}^c}D_{\mu}u_{R,s})$ &
      81 & & & \\
\hline 
      ${\cal O}_{HLQ^2uV}$ &  & 243 &
\ytableausetup{centertableaux}
3 \hskip-3mm \begin{ytableau}
       \none & s & \none \\
       \none & t & \none \\
\end{ytableau}
\hskip-3mm + \ytableausetup{centertableaux}
3 \hskip-3mm\begin{ytableau}
       \none & s & t& \none \\
\end{ytableau}
    & $e^+V$, $\pi^0e^+V$, & $\pi^-e^+ V$, \\
    & & & & $\pi^+\overline{\nu}V$ & 
      $\pi^0\overline{\nu}V$ \\
  1 &  $\epsilon^{\alpha\beta\gamma}\epsilon_{ik}\epsilon_{jl}
       (\overline{L_{p}^{i,c}}\gamma^{\mu}u_{R,r,\alpha})
        (\overline{Q_{t,\beta}^{j,c}}D_{\mu}Q^{k}_{s,\gamma})H^\dagger_l$ &
       & & & \\
  2 &  1: $ \epsilon_{ik}\epsilon_{jl} \to \epsilon_{ij}\epsilon_{kl}$ &
       & & & \\
  3 &  $\epsilon^{\alpha\beta\gamma}\epsilon_{ik}\epsilon_{jl}
      (\overline{L_{p}^{i,c}}Q^{k}_{s,\alpha})
        (\overline{Q_{t,\beta}^{j,c}}\gamma^{\mu}u_{R,r,\gamma}) D_{\mu}H^\dagger_l$ &
       & & & \\
  4 &  3: $ \epsilon_{ik}\epsilon_{jl} \to \epsilon_{ij}\epsilon_{kl}$ &
       & & & \\
  5 &  $\epsilon^{\alpha\beta\gamma}\epsilon_{ik}\epsilon_{jl}
      (\overline{L_{p}^{i,c}}Q^{j}_{t,\alpha})
        (\overline{Q_{s,\beta}^{k,c}}\gamma^{\mu}u_{R,r,\gamma}) D_{\mu}H^\dagger_l$ &
       & & & \\
  6 &  5: $ \epsilon_{ik}\epsilon_{jl} \to \epsilon_{ij}\epsilon_{kl}$ &
       & & & \\
\hline
      ${\cal O}_{HLQ^2dV}$ &  & 243 &
 \ytableausetup{centertableaux}
3 \hskip-3mm \begin{ytableau}
       \none & s & \none \\
       \none & t & \none \\
\end{ytableau}
\hskip-3mm + \ytableausetup{centertableaux}
3 \hskip-3mm\begin{ytableau}
       \none & s & t& \none \\
\end{ytableau}
    & $e^+V$, $\pi^0e^+V$, & $\pi^-e^+ V$, \\
    & & & & $\pi^+\overline{\nu}V$ & 
      $\pi^0\overline{\nu}V$ \\
  1 &  $\epsilon^{\alpha\beta\gamma}\epsilon_{ik}\epsilon_{jl}
       (\overline{L_{p}^{i,c}}\gamma^{\mu}d_{R,r,\alpha})
        (\overline{Q_{t,\beta}^{j,c}}D_{\mu}Q^{k}_{s,\gamma})H_l$ &
       & & & \\
  2 &  1: $ \epsilon_{ik}\epsilon_{jl} \to \epsilon_{ij}\epsilon_{kl}$ &
       & & & \\
  3 &  $\epsilon^{\alpha\beta\gamma}\epsilon_{ik}\epsilon_{jl}
      (\overline{L_{p}^{i,c}}Q^{k}_{s,\alpha})
        (\overline{Q_{t,\beta}^{j,c}}\gamma^{\mu}d_{R,r,\gamma}) D_{\mu}H_l$ &
       & & & \\
  4 &  3: $ \epsilon_{ik}\epsilon_{jl} \to \epsilon_{ij}\epsilon_{kl}$ &
       & & & \\
  5 &  $\epsilon^{\alpha\beta\gamma}\epsilon_{ik}\epsilon_{jl}
      (\overline{L_{p}^{i,c}}Q^{j}_{t,\alpha})
        (\overline{Q_{s,\beta}^{k,c}}\gamma^{\mu}d_{R,r,\gamma}) D_{\mu}H_l$ &
       & & & \\
  6 &  5: $ \epsilon_{ik}\epsilon_{jl} \to \epsilon_{ij}\epsilon_{kl}$ &
       & & & \\
\hline
      ${\cal O}_{HeQ^3V}$ &  & 81 &
 \ytableausetup{centertableaux}
\hskip-3mm\begin{ytableau}
       \none & r & \none \\
       \none & s & \none \\
       \none & t & \none \\
\end{ytableau}
\hskip-3mm + \hskip-3mm
\begin{ytableau}
       \none & r & s & t& \none \\
\end{ytableau}
& $e^+V$, $\pi^0e^+V$ & $\pi^-e^+ V$ \\
& & & \hskip-3mm+ 2 \hskip-3mm
\ytableausetup{centertableaux}
\begin{ytableau}
       \none & r & s & \none \\
       \none & t & \none \\
\end{ytableau}
& &  \\
  1 &  $\epsilon^{\alpha\beta\gamma}\epsilon_{ik}\epsilon_{jl}
       (\overline{Q_{r,\alpha}^{i,c}}\gamma^{\mu}e_{R,p})
        (\overline{Q_{s,\beta}^{j,c}}D_{\mu}Q^{k}_{t,\gamma})H_l$ &
       & & & \\
  2 &  1: $ \epsilon_{ik}\epsilon_{jl} \to \epsilon_{ij}\epsilon_{kl}$ &
       & & & \\
  3 &  $\epsilon^{\alpha\beta\gamma}\epsilon_{ik}\epsilon_{jl}
       (\overline{Q_{r,\alpha}^{i,c}}Q_{t,\beta}^{k,c})
        (\overline{Q_{s,\gamma}^{j,c}}\gamma^{\mu}e_{R,p}) D_{\mu}H_l$ &
       & & & \\
  4 &  3: $ \epsilon_{ik}\epsilon_{jl} \to \epsilon_{ij}\epsilon_{kl}$ &
       & & & \\
  5 &  $\epsilon^{\alpha\beta\gamma}\epsilon_{ik}\epsilon_{jl}
       (\overline{Q_{r,\alpha}^{i,c}}Q_{s,\beta}^{j,c})
        (\overline{Q_{t,\gamma}^{k,c}}\gamma^{\mu}e_{R,p}) D_{\mu}H_l$ &
       & & & \\
  6 &  5: $ \epsilon_{ik}\epsilon_{jl} \to \epsilon_{ij}\epsilon_{kl}$ &
       & & & \\
\hline
\end{tabular}
\caption{
SMEFT operators containing a single  gauge vector  at $d=8$ that
violate baryon and lepton number and can induce nucleon decay. The
notation and conventions are the same as in
Tab.~\ref{tab:Opd7}. All operators shown are of the type
$\psi^4 D H$. For the last three operator classes the total number of
independent operator coefficients and the corresponding flavour
symmetry properties are indicated. In these cases, the operators
labeled by (1)-(6) do not individually possess definite symmetry
properties and must be combined appropriately to form irreducible
flavour structures. The corresponding combinations are listed in
Tab.~\ref{tab:sym8}. }
\label{tab:Opd8}
\end{table}

At $d=8$, multiple Lorentz and, in several cases, multiple
$\mathrm{SU}(2)$ contractions can lead to independent operator
structures. For operators involving only two $\mathrm{SU}(2)$
doublets, the symmetric and antisymmetric combinations can be defined
straightforwardly and are given directly in
Tab.~\ref{tab:Opd8}. For operators involving four
$\mathrm{SU}(2)$ doublets, the construction of operators with definite
symmetry properties is more involved. In these cases,
in Tab.~\ref{tab:Opd8} we list only the independent operator structures,
while in Tab.~\ref{tab:sym8} we provide the combinations with definite
symmetry properties. For these combinations we directly adopt the
output convention of \texttt{AutoEFT}~\cite{Harlander:2023ozs}.

\begin{table}[h]
\renewcommand*{\arraystretch}{1.0}
\centering
\ytableausetup
{boxsize=0.8em}
\begin{tabular}{|c|c|c|c|}
  \hline
 Type & Combination & Parameters & Symmetry \\
\hline
 $Q^2$ & ${\cal O}_3 + {\cal O}_4 - {\cal O}_5 - {\cal O}_6$ & 27 &
 \ytableausetup{centertableaux}
\hskip-3mm\begin{ytableau}
       \none & s & \none \\
       \none & t & \none \\
\end{ytableau}
\\
& ${\cal O}_3 - {\cal O}_4 + {\cal O}_5 - {\cal O}_6$ & 27 &
 \ytableausetup{centertableaux}
\hskip-3mm\begin{ytableau}
       \none & s & \none \\
       \none & t & \none \\
\end{ytableau}
\\
& ${\cal O}_3 - {\cal O}_4 - {\cal O}_5 + {\cal O}_6$ & 54 &
 \ytableausetup{centertableaux}
\hskip-3mm\begin{ytableau}
       \none &s & t & \none \\
\end{ytableau}
\\
& ${\cal O}_3 + {\cal O}_4 + {\cal O}_5 + {\cal O}_6$ & 54 &
 \ytableausetup{centertableaux}
\hskip-3mm\begin{ytableau}
       \none & s & t & \none \\
\end{ytableau}
\\
 & $2 ({\cal O}_1 - {\cal O}_2)
      - {\cal O}_3 + {\cal O}_4 + {\cal O}_5 - {\cal O}_6 $ & 27 &
 \ytableausetup{centertableaux}
\hskip-3mm\begin{ytableau}
       \none & s & \none \\
       \none & t & \none \\
\end{ytableau}
\\
 & $2 ({\cal O}_1 + {\cal O}_2)
      - {\cal O}_3 - {\cal O}_4 + {\cal O}_5 + {\cal O}_6$ & 54 &
\hskip-3mm\begin{ytableau}
       \none & s & t & \none \\
\end{ytableau}
\\
\hline
 $Q^3$ & ${\cal O}_4 - {\cal O}_5$ & 3 &
 \ytableausetup{centertableaux}
\hskip-3mm\begin{ytableau}
       \none & r & \none \\
       \none & s & \none \\
       \none & t & \none \\
\end{ytableau}
\\
 & $2 {\cal O}_3 - {\cal O}_4 - {\cal O}_5 - {\cal O}_6$ & 24 &
 \ytableausetup{centertableaux}
\hskip-3mm\begin{ytableau}
       \none & r & s &\none \\
       \none & t & \none \\
\end{ytableau}
\\
 & $-2 {\cal O}_3 + {\cal O}_4 + {\cal O}_5 - 2 {\cal O}_6$ & 30 &
 \ytableausetup{centertableaux}
\hskip-3mm\begin{ytableau}
       \none & r & s & t & \none \\
\end{ytableau}
\\
 & $6 {\cal O}_1 - 3 {\cal O}_2 - 4 {\cal O}_3 + 2 {\cal O}_4
  + 2 {\cal O}_5 - {\cal O}_6$ & 24 &
 \ytableausetup{centertableaux}
\hskip-3mm\begin{ytableau}
       \none & r & s &\none \\
       \none & t & \none \\
\end{ytableau}
\\
\hline
\end{tabular}
\caption{Combinations of operator structures from Tab.~\ref{tab:Opd8} with
four $\mathrm{SU}(2)$ doublets that possess definite flavour symmetry
properties.  The ``Combination'' column lists the
linear combinations of operators that have definite symmetries and gives the number of parameters contained. Operator classes shown: (i)
$Q^2 \rightarrow {\cal O}_{HLQ^2uV}, {\cal O}_{HLQ^2dV}$ and (ii)
$Q^3 \rightarrow {\cal O}_{HeQ^3V}$. In the latter case, only four of
the six operator structures listed in Tab.~\ref{tab:Opd8} are
linearly independent.}
\label{tab:sym8}
\end{table}

We briefly comment on the differences
between operator constructions involving  gauge  vectors and those
in which vector states are introduced explicitly into the field
content. From the EFT point of view, operators
containing an explicit vector field can be constructed solely by
requiring Lorentz invariance, without specifying the microscopic
origin of the vector field\footnote{For example, the vector could arise as
a remnant of some
strongly interacting confined sector.}. The resulting
operator basis, however, is not equivalent to that obtained for gauge
vectors. Both the parameter counting and the symmetry properties of
the operators generally differ. Using
\texttt{Sym2Int}~\cite{Fonseca:2017lem}, we have also constructed the
corresponding operator basis for explicit non-gauge vectors.

Although the Lorentz structures of the operators are identical in the
two cases described, the counting of independent operators differs. For gauge
vectors, the vector field appears through the covariant derivative 
and equations of motion  together with integration by parts
 identities can be used to relate different operator structures.
This frequently reduces the number of independent operators and hence
the number of resulting free parameters. Conversely, when the covariant
derivative acts on fermion bilinears, the precise placement of the
derivative becomes physically relevant. This distinction does not
arise for explicit non-gauge vectors. Consequently, certain operator
classes can contain either fewer or more independent structures in the
gauge vector case relative to the explicit vector case.

The symmetry properties of some operators also differ between the
gauge vector and non-gauge vector constructions. A prominent example is the $d=7$ operator ${\cal O}_{ed^3V}$. As shown
in Tab.~\ref{tab:Opd7}, for a gauge vector this operator is symmetric
in the flavour indices $(r,s,t)$. In contrast, for an explicit
non-gauge vector the operator must be completely antisymmetric in
$(r,s,t)$ and therefore vanishes in the single generation limit for
down-type quarks. As a consequence, the gauge vector realization of
${\cal O}_{ed^3V}$ can induce both $\pi^+ e^- V$ and
$K^+ e^- V$ final states, whereas the non-gauge vector case allows
only the $K^+ e^- V$ channel. We note that ${\cal O}_{ed^3V}$ is the
only operator in our operator basis that vanishes in the
single generation limit for explicit non-gauge vectors.

\subsection{B. Ultraviolet Completion \label{app:UV}}

We comment on possible ultraviolet (UV) completions and the broader
implications for light vector new physics.  While the light massive
vector can be treated phenomenologically as a generic spin-1 field at
low energies, generic Proca interactions imply a finite UV cutoff due
to the growth of amplitudes involving longitudinal vector
polarizations. A controlled UV completion can be obtained if the
vector originates from a gauge symmetry, with the mass generated
through either a Higgs or Stueckelberg mechanism. We focus on
realizations in which the vector originates from a spontaneously
broken gauge symmetry, thus avoiding these
issues\footnote{Alternatively, the vector could arise as a composite
  state of a strongly coupled sector, with the strong dynamics
  unitarizing the UV behaviour. The effective description considered
  here would then remain valid below the confinement scale.}.  For the
operator classes listed in Sec.~A, the vector field can arise either
through the covariant derivative or through the corresponding gauge
field strength tensor. Up to $d = 8$, the lowest dimensional baryon
number-violating operators arise from the covariant derivative
structure\footnote{Since the field strength tensor has mass dimension
  two, operators involving field strengths appear at higher dimension
  than those constructed from covariant derivatives.}.

Existing constraints on light vectors from searches such as from beam
dumps, together with the requirement of macroscopic decay lengths of
vectors, motivate weakly coupled realizations with couplings
significantly smaller than electroweak gauge couplings. If the vector
arises from a broken local gauge symmetry, it is of interest to
consider how its light mass and suppressed couplings can appear in the
UV dynamics.

To illustrate the relevant realization in the UV setting, consider a simple $\text{U}(1)'$
extension of the SM allowing both direct gauge couplings to SM
fermions and kinetic mixing with SM hypercharge. In the unbroken phase,
the relevant gauge-sector Lagrangian, including mixing between the
hypercharge boson $B$ and the $\text{U}(1)'$ boson $B^\prime$, is given by
\begin{equation}{\label{eq:gkl1}}
  \mathcal L^\text{gauge}\supset
  -\frac{1}{4}\tilde{B}_{\mu\nu} \tilde{B}^{\mu\nu}
  - \frac{1}{4} \tilde{B^\prime}_{\mu\nu}\tilde{B^\prime}^{\mu\nu}
  + \frac{\epsilon_k}{2} \tilde{B}_{\mu\nu}\tilde{B^\prime}^{\mu\nu} +
  \sum_{f} i \bar f  \tilde{\slashed{D}}  f  ,
\end{equation}
where, $\tilde{B}_{\mu\nu}$ and $\tilde{B^\prime}_{\mu\nu}$
are the associated respective field strengths and
$\epsilon_k$ is the kinetic mixing parameter.
The gauge covariant derivative can be expressed as
\begin{equation}{\label{eq:gkl2}}
  \tilde{D}_\mu  =   \partial_\mu + \dots + i  g_1   Y_f  
  \tilde{B}_\mu
  + i  g'   Q_f^{X}   \tilde{B}^\prime_\mu  ,
\end{equation}
where the hypercharge and $\text{U}(1)'$ charge are denoted by $Y_f =
Q_f-T_{3 f}$ and $Q^{X}_f$, respectively, with the corresponding
gauge couplings denoted by $g_1$ and $g'$.  The gauge kinetic terms in
Eq. \eqref{eq:gkl1} can then be brought to the diagonal canonical form by
employing a linear transformation of the fields, $\tilde{B}_\mu =
B_\mu +  \epsilon_k B^\prime_\mu/(\sqrt{1 - \epsilon_k^2})$
and $\tilde{B^\prime}_\mu = B^\prime_\mu/(\sqrt{1-\epsilon_k^2})$.  Following the gauge symmetry breaking, the physical
neutral gauge boson mixing can be written as
 \begin{equation}
  \begin{pmatrix}A^\mu\\ Z^\mu\\X^{\mu}\end{pmatrix}
  = \begin{pmatrix}
  \cos\theta_w & \sin\theta_w & 0\\
  - \sin\theta_w \cos\theta^\prime & \cos\theta_w\cos\theta^\prime &
  \sin\theta^\prime\\ 
  \sin\theta_w\sin\theta^\prime & -\cos\theta_w\sin\theta^\prime &
  \cos\theta^\prime 
  \end{pmatrix}
  \begin{pmatrix}B^\mu\\
    W_3^\mu\\B^{\prime \mu}\end{pmatrix} ,
 \end{equation}
where $\theta_w$ is the standard weak mixing angle and $\theta'$ is
given by 
\begin{equation}
\tan2\theta^\prime  =  \dfrac{2 \varepsilon^\prime  g_1 
  \sqrt{g^2 + {g_1}^2}}{{ {\varepsilon^\prime}^2 {g_1}^2 + 4 
(m_{B^\prime}^2/v^2) - g^2 - {g_1}^2}} ,
\end{equation}
keeping only linear order terms in $\varepsilon^\prime$, where
$\varepsilon^\prime ={\epsilon_k}/{\sqrt{1-\epsilon_k^2}}$. In the small kinetic-mixing limit,
$\epsilon_k \ll 1$, the mass eigenvalue of the neutral vector bosons is approximately
$m_X \simeq m_B'=2 v_X g_X$, where $v_X$ denotes the symmetry-breaking
vacuum expectation value associated with $\text{U}(1)_X$.
The relevant terms of the gauge covariant derivative can thus
be expressed as\footnote{Here, we neglect higher order corrections in
the $Z$ coupling due to mixing with the $X$ at order
${\varepsilon^\prime}^2\:\text{or}\:\varepsilon^\prime g^\prime$.}
\begin{equation}\label{eq:CDE}
  D_\mu  \simeq  \partial_\mu + \dots + i \frac{g}{\cos\theta_w} 
  (T_{3 f} - \sin^2\theta_W  Q_f) Z_\mu + ie  Q_f  A_\mu +
  i (e \varepsilon 
  Q_f + \varepsilon_{X} Q_f^{X}) X_\mu  ,
\end{equation}
with $\varepsilon  =   {\epsilon_k \cos\theta_w}/
\left({\sqrt{1-\epsilon_k^2}} \right)$ and $\varepsilon_{X}
 = {g_{X}}/\left({ \sqrt{1 -\epsilon_k^2}}\right)$.

Equation~\eqref{eq:CDE} shows that the light vector can decay into SM
states either through kinetic mixing, analogously to the standard dark
photon scenario via $\varepsilon$, or through direct couplings associated with the new
$\text{U}(1)_X$ gauge symmetry $\varepsilon_X$.  Realizing a
sub-GeV vector with $v_X$ significantly above the electroweak scale
therefore requires a suppressed gauge coupling $g_X \ll 1$. If the extended new $\text{U}(1)$ emerges from a single large GUT
gauge group together with the SM gauge group, then $g_X\ll1$ must be
realized via strong renormalization group  running, starting from the usual
$\mathcal{O}(1)$ coupling at GUT scale. A particularly simple and technically
natural alternative is provided by sequestered or hidden-sector
realizations, in which the light vector originates from an additional
weakly coupled sector largely decoupled from the GUT dynamics.

\begin{figure*}[t]
  \begin{center}
    \includegraphics{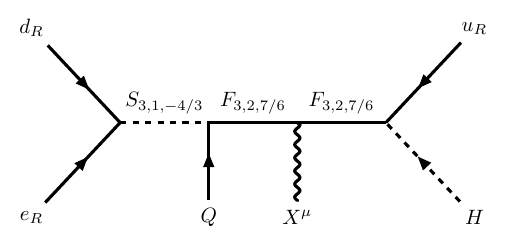}\hfill
    \includegraphics{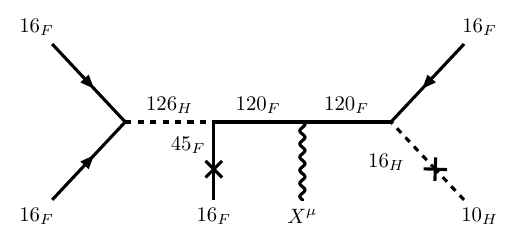}
    \caption{Illustration of the representative UV completion discussed in the
text. (Left) Example tree-level decomposition of the effective operator
${\cal O}_{HeQduV}$ in the SM phase. (Right) Corresponding realization within the
$\text{SO}(10)\times \text{U}(1)_X$ framework.}
    \label{fig:UV_ex}
\end{center}
\end{figure*}

To illustrate explicitly how the effective interactions discussed in
this work can arise from a UV-complete theory, let us consider the
$d = 8$ operator
${\cal O}_{HeQduV}\equiv
H(\overline{Q^{c}}\gamma^{\mu}e_{R})
(\overline{u_{R}^c}D_{\mu}d_{R})$
from Tab.~\ref{tab:Opd8}. A simple representation analysis shows
that this operator can be generated at tree level through several
distinct heavy mediator topologies. One representative realization is
shown in Fig.~\ref{fig:UV_ex}. The model contains two heavy states, a
vector-like coloured electroweak doublet fermion
$F:(3,2,7/6,q_X^F)$ and a coloured scalar
$S:(3,1,-4/3,q_X^S)$, where the entries denote the transformation
properties under
$(\text{SU}(3)_C,\text{SU}(2)_L,\text{U}(1)_Y,\text{U}(1)_X)$. Gauge invariance under $\text{U}(1)_X$ implies the charge relations
\begin{equation}
q_X^S
=
q_X^{e_R}+q_X^{d_R}
=
-q_X^Q-q_X^F
=
-q_X^Q-q_X^u ,
\end{equation}
where the SM Higgs doublet is assumed to be neutral under
$\text{U}(1)_X$. Consequently, all fields carrying non-vanishing $\text{U}(1)_X$
charge can emit the light vector boson $V$, and all corresponding
contributions must be included in the full effective interaction.

As an illustrative UV completion, we consider an embedding of this
field content into a sequestered
$\text{SO}(10)\times \text{U}(1)_X$ framework. Owing to the sequestered nature of
the additional $\text{U}(1)_X$ sector, the associated gauge coupling $g_X$
and the mass generation mechanism for the light vector are largely
decoupled from the $\text{SO}(10)$ dynamics and proceed as discussed above.
It may be of interest to consider how the required field content, interactions, and relevant mass scales can arise from a
UV-complete $\text{SO}(10)$ framework. While a detailed study of a full GUT
model is beyond the scope of this work, we discuss, in what follows,
an example realization that highlights some of the key insights applicable to a wide
variety of UV-complete scenarios.

Let us consider the standard embedding of the SM chiral fermions into
the $16_{F_i}$ representations of $\text{SO}(10)$, where
$i=1,2,3$ denotes the family index, and assume the intermediate
symmetry breaking pattern\footnote{A linear
combination of $\text{U}(1)_\eta$ and $\text{U}(1)_X$ can also be considered.
Here, for simplicity, we assume that the light vector boson originates
entirely from the sequestered $\text{U}(1)_X$ sector.}
$\text{SO}(10)\rightarrow \text{SU}(5)\times \text{U}(1)_\eta$.
Under $\text{SU}(5)\times \text{U}(1)_\eta$, the decomposition of the
$16_F$ representation is
\begin{equation}
16_F \rightarrow 10(1)+\bar{5}(-3)+1(5) ,
\end{equation}
where the numbers in parentheses denote the corresponding
$\text{U}(1)_\eta$ charges, normalized conventionally by
$1/(2\sqrt{10})$.
The $\text{SU}(5)$ multiplets $10$ and $\bar{5}$ contain the usual SM field
content,
\begin{equation}
10 \supset
Q:(3,2,1/6) ,\quad
u^c:(\bar{3},1,-2/3) ,\quad
e^c:(1,1,1) ,
\end{equation}
and
\begin{equation}
\bar{5}\supset
d^c:(\bar{3},1,1/3) ,\quad
L:(1,2,-1/2) ,
\end{equation}
where we use standard left chiral notation. The Higgs representation
$10_H$ contains the SM Higgs doublet and generates fermion masses
through Yukawa interactions of the form
$16_F 16_F 10_H$.

The coloured scalar
$S:(3,1,-4/3)$ can be embedded into the
$126_H$ representation of $\text{SO}(10)$, which decomposes as
\begin{equation}
126_H
\rightarrow
50(-2)+\overline{45}(2)+15(6)
+\overline{10}(-6)+5(2)+1(10) ,
\end{equation}
with the $\overline{45}$ of $\text{SU}(5)$ containing the scalar $S$. This
embedding allows the interaction vertex involving
$S$, $e_R$, and $d_R$ to arise from Yukawa interactions of the form
$16_F 16_F 126_H$. Embedding
$S:(3,1,-4/3)$ into $120_H$ is less suitable because it would require
generation-antisymmetric couplings of the two $16_F$ multiplets.
Moreover, the $120_H$ representation generically induces both
leptoquark-type  and diquark-type couplings~\cite{Aulakh:2006hs,Nath:2001uw},
potentially generating lower dimensional operators leading to rapid proton decay. In contrast, the $126_H$ embedding considered here permits
only leptoquark-type interactions for $S$, thereby avoiding these
operators and allowing comparatively low scalar masses in the presence
of multiplet splitting.

The vector-like electroweak doublet fermion
$F:(3,2,7/6)$ can arise from the real representation $120_F$, with
decomposition
\begin{equation}
120_F
\rightarrow
\overline{45}(2)+45(-2)+\overline{10}(-6)
+10(6)+\overline{5}(2)+1(10) ,
\end{equation}
where the $45+\overline{45}$ pair contains the vector-like fermion
$F$, while the remaining multiplets acquire masses near the GUT
scale.

The interaction involving
$S\subset126_H$,
$F\subset120_F$, and
$Q\subset16_F$
is not an $\text{SO}(10)$ singlet at tree level. However, it can be generated
through mixing of $Q$ with a vector-like partner
$Q'\subset45_F$~\cite{King:2017anf}. Under
$\text{SU}(5)\times \text{U}(1)_\eta$, the $45_F$ decomposes as
\begin{equation}
45_F
\rightarrow
24(0)+\overline{10}(4)+10(-4)+1(0) ,
\end{equation}
with the $10+\overline{10}$ pair containing the vector-like state
$Q'$. The required mixing can then be induced through Higgs
representations
$16_H+\overline{16}_H$,
allowing Yukawa interactions of the form
$16_F 45_F \overline{16}_H$.

Mixing between $16_H$ and $10_H$ induced by the vacuum expectation
value of $16_H$, below the $\text{U}(1)_\eta$ breaking scale but above the
electroweak scale, can additionally assist in realizing
doublet-triplet splitting~\cite{Antusch:2017tud}. The same mixing also
generates the remaining interaction vertex involving
$F\subset120_F$,
$u_R\subset16_F$,
and the SM Higgs field. Finally, since the fermionic multiplets
introduced to realize the vector-like fermions are real
representations, they do not induce gauge anomalies in the
$\text{SO}(10)$ theory.

\subsection{C. Nucleon Decay Rate Calculation \label{app:kin}} 

In the absence of the ab initio computations for the energy-dependent nuclear matrix elements for the nucleon decay modes of our interest, which involve quark currents involving Lorentz vector structure, matching our SM gauge group invariant effective operators to low energy EFT and subsequently to chiral perturbation theory (ChPT) provides a systematic way to estimate the relevant decay rates. 

The  low energy constants can be estimated by employing power counting in  dimensional analysis~\cite{Weinberg:1989dx,Manohar:1983md}. Recent relevant ChPT-based prescriptions~\cite{Liao:2025vlj,Song:2026gyo} have been put forth for all allowed chiral structures under the QCD chiral group $\text{SU}(3)_L\times \text{SU}(3)_R$ for matching baryon number violating interactions involving light new physics states. Considering an illustrative two-body decay mode  $p(p)\rightarrow e^+ (k) V(q)$, the SM gauge group invariant operator $\mathcal{O}_{Heu^2Q}$  discussed in the main text can be matched onto the leading order $\text{SU}(3)_L\times \text{SU}(3)_R$ invariant chiral term of the form
\begin{equation}{\label{eq:cpt1}}
\mathcal{L} \supset {c \over \Lambda_\chi} 
 {\cal P}_{yzi}^{RL,\mu}
\left(g_{\mu\nu} - {1\over 4} \gamma_\mu \gamma_\nu\right)\text{P}_{R}
(\xi^\dagger i D^\nu \text{P}_{R} B  \xi^\dagger)_{yj}
\Sigma^*_{kz} \epsilon_{ijk}  \supset  {c \over \Lambda_\chi} {\cal P}_{uud}^{RL,\mu} i \left(\partial_\mu-{1 \over 4} \gamma_\mu \slashed{\partial}\right)\text{P}_{R}   p(p),
\end{equation}
where we follow the notation and chiral building blocks of Ref.~\cite{Liao:2025vlj}, denoting the baryon octet matrix by $B(x)$ and $\Sigma(x)=\xi^2=\exp\left( i\sqrt{2} \Pi(x)/F_0\right)$, where $\Pi(x)$ is the meson octet. The indices $y,z,i,j,k$ run over the light quark flavors $u,d,s$. Here, $g_{\mu\nu}$, $\gamma_{\mu}$  and $\text{P}_{R}$ denote the metric tensor, Dirac gamma matrix  and projection operators, respectively. The second expression to the right in Eq.~\eqref{eq:cpt1} is obtained by expanding the matrices to keep terms only up to zeroth order in the pseudoscalar meson field. The spurion field ${\cal P}_{yzi}^{RL,\mu}$ contains all the non-quark fields that transform trivially under the QCD chiral group. The Wilson coefficient obtained by integrating out the heavy new physics fields in the high-scale EFT, in this case, is given by ${\cal P}_{yzi}^{RL,\mu}= C_{Heu^2Q} (\Lambda_\chi) v  \overline{e^c} V^\mu / (\sqrt{2} \Lambda^4)$,  where $C_{Heu^2Q}(\Lambda_\chi)$ is obtained by including any renormalization group running effects between the scales $\Lambda$ and $\Lambda_\chi$. 

Using Eq.~\eqref{eq:cpt1} and the expression for the spurion field above, the
amplitude at the chiral level can then be expressed in the form 
\begin{align}{\label{eq:cpt2}}
\mathcal{M}_{p\to e^+ V}^{\text{ChPT}}
&=
\left(\frac{c  C_{Heu^2Q} v }{\sqrt{2} \Lambda_\chi  \Lambda^4} \right)
\bar u_e (p_\mu-{1 \over 4} \gamma_\mu \slashed{p}) u_p 
\epsilon^{*\mu}
\end{align}
where $u$ with subscripts $e$ and $p$ dotes the corresponding spinors and $\epsilon^{*\mu}$ is the polarization vector of the vector field. Eq.~\eqref{eq:cpt1} can be expressed in the form given in Eq.~\eqref{eq:FF2b} using Dirac algebra computations, orthogonality of the polarization vector to its momentum  and momentum conservation relations allowing to obtain the expressions for the form factors given in the main text. Finally, the low energy constant $c$ can be determined by matching the effective reduced couplings for the quark- and hadron-level operators at the chiral scale. In the dimensional analysis~\cite{Weinberg:1989dx,Manohar:1983md}, the effective reduced coupling is defined as $g (4\pi)^{2-m}\Lambda_\chi^{d-4}$, where $g$ is the coupling constant and $m$ is the number of physical fields contained in a $d$-dimensional operator. From the final form in Eq.~\eqref{eq:cpt1}, the hadron-level interaction part $(c/\Lambda_\chi) (\partial_\mu-{1 \over 4} \gamma_\mu \slashed{\partial})   p$ gives $g=c/\Lambda_\chi$, $m=1$ and $d=5/2$, leading to $C_H^{\text{ERC}}=  4\pi c \Lambda_\chi^{-5/2}$. On the other hand, at the quark level from Eq.~\eqref{eq:EFO} we have $g=1$, $m=3$ and $d=9/2$, leading to $C_q^{\text{ERC}}=(4\pi)^{-1} \Lambda_\chi^{1/2}$. Equating $C_H^{\text{ERC}}$ and $C_q^{\text{ERC}}$ yields $c\simeq \Lambda_\chi^3/(4\pi)^2$.

Following discussion for the two-body decay case in the main text, the amplitude for the three-body decay $p (p)\rightarrow  \pi^0 (p') e^+ (k) V(q')$ process depicted in Fig.~\ref{fig:DblBng}  can analogously be expressed in terms of the independent Lorentz structures
\begin{equation}{\label{eq:FF3b}}
 {\mathcal{M'}}= \bar{u}_{e} \text{P}_{R} \left[ G_1(q^2) \gamma^{\mu}  + G_2(q^2 ) i \sigma^{\mu\nu} q_\nu  + G_3(q^2 )  q^\mu \right] u_{p} \epsilon^{\ast}_{\mu}~,
\end{equation}
where $q=p-k=p'+q'$. Note that, due to the increased number of independent external momenta, the number of independent Lorentz structures has now increased relative to the two-body case. Here, we will also consider the same example operator $\mathcal{O}_{Heu^2Q}$ as the origin of this decay mode. In this case also, a contact interaction arising from the same $\text{SU}(3)_L\times \text{SU}(3)_R$ invariant leading order chiral term can contribute to the amplitude. Expanding it, we obtain the linear order terms in the pseudoscalar meson field
\begin{equation}{\label{eq:cpt3}}
\mathcal{L} \supset {c \over \Lambda_\chi} 
 {\cal P}_{yzi}^{RL,\mu}
\left(g_{\mu\nu} - {1\over 4} \gamma_\mu \gamma_\nu\right)\text{P}_{R}
(\xi^\dagger i D^\nu \text{P}_{R} B  \xi^\dagger)_{yj}
\Sigma^*_{kz} \epsilon_{ijk}  \supset  {i \over F_0}{3 \over 2}{c \over \Lambda_\chi} {\cal P}_{uud}^{RL,\mu} \pi^0 \left(\partial_\mu-{1 \over 4} \gamma_\mu \slashed{\partial}\right)\text{P}_{R}   p(p)  ,
\end{equation}
where $F_0=82.2\pm 0.5 \text{ MeV}$ is the pion decay constant. In addition to the contact interaction above, the decay amplitude also receives another contribution due to the exchange of a decuplet baryon in the leading order chiral Lagrangian. While we have included this contribution explicitly in the numerical estimates of the relevant decay width (see e.g. Ref.~\cite{Liao:2025vlj} for an explicit example of a similar computation), we find that its contributions are subdominant giving a correction of less than $\mathcal{O}(10)\%$ to the contact interaction  and, for simplicity, we will ignore it in the current discussion. 

The short-range contribution to the three-body decay $p (p)\rightarrow  \pi^0 (p') e^+ (k) V(q')$ amplitude at the chiral level can then be expressed in the form 
\begin{align}{\label{eq:cpt5}}
\mathcal{M}_{p\to e^+\pi^0 V}^{\text{ChPT}}
&=
\left(\frac{3 c  C_{Heu^2Q} v }{2\sqrt{2} F_0 \Lambda_\chi  \Lambda^4} \right)
\bar u_e (p_\mu-{1 \over 4} \gamma_\mu \slashed{p}) u_p 
\epsilon^{*\mu}~,
\end{align}
where the momentum $p$ is now $p=k+p'+q'$. Following similar approach as in the two-body decay case Eq.~\eqref{eq:cpt5} can be expressed in the form similar to Eq.~\eqref{eq:FF3b} allowing to determine the relevant form factors $G_1=C'(\Lambda, \Lambda_\chi) m_{p}/4$, $G_2=C'(\Lambda, \Lambda_\chi)/2=G_3$, with $C'(\Lambda,
\Lambda_\chi)=
3 c(\Lambda_\chi) v C_{Heu^2Q} / (2 \sqrt{2} F_0 \Lambda_\chi \Lambda^4 )$, where we have again neglected the light SM lepton masses. 

Once we have the amplitudes as described above, the computation for the decay widths follows the standard prescription as we detail here. For a two-body decay $X(p,m) \to A(p_1,m_1) B(p_2,m_2)$ in the
center-of-mass frame, the decay width is~\cite{ParticleDataGroup:2024cfk} 
\begin{equation}
    \Gamma_\text{2B} = \frac{1}{16 \pi} \frac{\sqrt{\lambda(m^2,m_1^2,m_2^2)}}{m^3} \vert\overline{\mathcal{M}_\text{2B}} \vert^2   ,
\end{equation}
where
\begin{equation}
    \lambda(x,y,z)=x^2+y^2+z^2-2xy-2xz-2yz
\end{equation}
is the Källen function and the bar denotes spin averaging over initial states and summing over final states. The magnitude of the final-state three momentum is
\begin{equation}
    |\vec p |
    =
    \frac{\sqrt{\lambda(m^2,m_1^2,m_2^2)}}{2m}.
\end{equation}
For a fixed vector mass $m_V$, this fixes the boost factor $\beta\gamma=|\vec p |/m_V$ in the free proton two-body decay limit.

\begin{figure}[t]
\centering
    \begin{subfigure}{0.49\linewidth}
        \centering
        \includegraphics{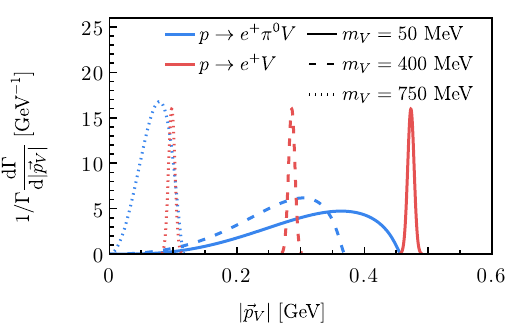} 
    \end{subfigure} \hfill
    \begin{subfigure}{0.49\linewidth}
        \centering
        \includegraphics{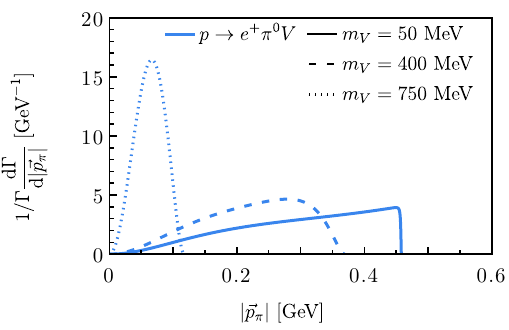}
    \end{subfigure}
    \medskip
    \begin{subfigure}{0.49\linewidth}
        \centering
        \includegraphics{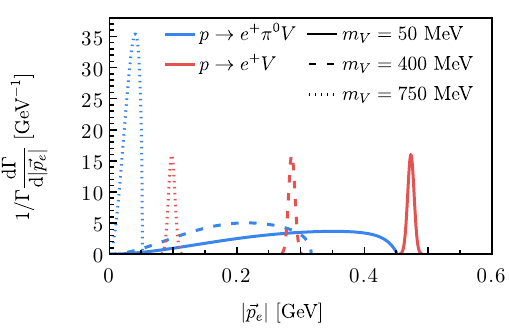} 
    \end{subfigure}\hfill
 \centering
    \caption{ Momentum distributions
    for free two-body and three-body nucleon decays into a vector boson $V$ induced
    by the operator $\mathcal{O}_{\bar{H}uuQeV}$. The curves are normalized by momentum integration to unity. Results are shown for masses
    $m_V = 50$, $400$, and $750~\mathrm{MeV}$, where kinematically allowed.}
    \label{fig:kinematics}
\end{figure}

For a three-body decay $X(p,m)\to A(p_1,m_1)B(p_2,m_2)C(p_3,m_3)$, we use the invariant mass variables
$m_{ij}^2=(p_i+p_j)^2$. The differential width is~\cite{ParticleDataGroup:2024cfk}
\begin{equation}
    \mathrm{d}\Gamma_{3{\rm B}}
    =
    \frac{1}{(2\pi)^3}
    \frac{1}{32m^3}
    \overline{|\mathcal{M}_{3{\rm B}}|^2}
     \mathrm{d}m_{12}^2 \mathrm{d}m_{23}^2 .
    \label{eq:dGdm}
\end{equation}

The decay width is obtained by integrating Eq.~\eqref{eq:dGdm}
\begin{equation}
\Gamma_{3{\rm B}}
=
\int_{(m_1+m_2)^2}^{(m-m_3)^2}
\mathrm{d}m_{12}^2
\int_{(m_{23}^2)_{\rm min}}^{(m_{23}^2)_{\rm max}}
\mathrm{d}m_{23}^2 
\frac{1}{(2\pi)^3}
\frac{1}{32m^3}
\overline{|\mathcal{M}_{3{\rm B}}|^2},
\end{equation}
with
\begin{align}
(m_{23}^2)_{\rm max/min}
&=
(E_2^*+E_3^*)^2
-
\left(
\sqrt{(E_2^*)^2-m_2^2}
\mp
\sqrt{(E_3^*)^2-m_3^2}
\right)^2,
\\
E_2^*
&=
\frac{m_{12}^2-m_1^2+m_2^2}{2m_{12}},
\\
E_3^*
&=
\frac{m^2-m_{12}^2-m_3^2}{2m_{12}} .
\end{align}
Here $E_2^*$ and $E_3^*$ are evaluated in the rest frame of the particle $(1,2)$ system.

Momentum distributions are obtained by changing variables. For example, for particle $C$ in the center of mass rest frame,
\begin{equation}
    m_{12}^2
    =
    m^2+m_3^2
    -
    2m\sqrt{m_3^2+|\vec p_3|^2},
\end{equation}
and therefore
\begin{equation}
    \frac{\mathrm{d}\Gamma}{\mathrm{d}|\vec p_3|}
    =
    \left|
    \frac{\mathrm{d}m_{12}^2}{\mathrm{d}|\vec p_3|}
    \right|
    \frac{\mathrm{d}\Gamma}{\mathrm{d}m_{12}^2}
    =
    \frac{2m|\vec p_3|}
    {\sqrt{m_3^2+|\vec p_3|^2}}
    \frac{\mathrm{d}\Gamma}{\mathrm{d}m_{12}^2}.
\end{equation}
The absolute value appears due to the Jacobian of the variable transformation  since $m_{12}^2$ decreases monotonically with $|\vec p_3|$. The resulting momentum distributions for representative vector masses are shown in Fig.~\ref{fig:kinematics}. In contrast to the two-body case, the three-body decay produces a continuous and generally non-Gaussian $\beta\gamma$ distribution. 
We note that the above analysis applies to free nucleon decays. For bound
nucleons in nuclei, these kinematics are modified by nuclear effects as
discussed below.

\begin{figure}[t]
    \centering \includegraphics[scale=1]{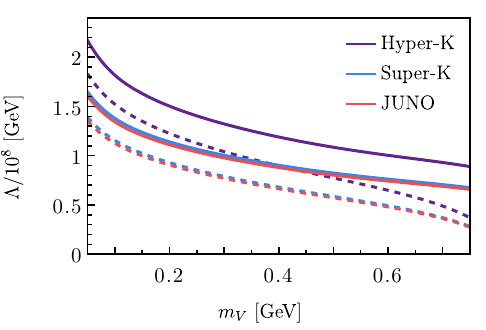}
 \caption{Sensitivity to the effective energy scale $\Lambda$ for the two-body
 (solid) and three-body (dashed) decay considering the operator
 $\mathcal{O}_{Heu^2Q}$. For the vertex resolution we use $l_{\min}=0.3~\mathrm{m}$ for Hyper-K and Super-K and $l_{\min}=0.6~\mathrm{m}$
 for JUNO. We
 assume 10 years of detector livetime for all experiments. We assume the Wilson
 coefficient, as well as detector efficiency are equal to unity. We consider
 $\beta \gamma c \tau = 3.5$~m.
\label{fig:Lambda}}
 
 \end{figure}

 For completeness, we present an alternative interpretation of the sensitivity compared to the lifetime shown in main text in terms of the effective scale of new physics beyond SM $\Lambda$ that can be probed and associated with   higher dimensional operators. The resulting reach in $\Lambda$ as a function of the vector mass $m_V$ is shown in Fig.~\ref{fig:Lambda} for representative detector exposures.

The translation from lifetime sensitivity to $\Lambda$ depends on the assumed operator structure and numerical prefactors. In particular, for the example $d=8$ operator considered here, the decay width scales as $\Gamma \propto \Lambda^{-8}$, such that the inferred reach in $\Lambda$ is sensitive to $\mathcal{O}(1)$ Wilson coefficients and hadronic matrix elements. Therefore, the interpretation in terms of $\Lambda$ is inherently model dependent and should be regarded as illustrative.

With these considerations, Fig.~\ref{fig:Lambda} indicates that current and upcoming large volume detectors can probe effective scales up to $\Lambda \sim \mathcal{O}(10^8) $GeV, with the exact reach depending on factors such as the detector exposure. While complementary, we emphasize that the lifetime projections presented in the main text provide a more model  independent characterization of the experimental sensitivity.

\subsection{D. Geometric Acceptance\label{app:geom}} 

Sensitivity to the displaced echo topology requires that both the primary nucleon decay vertex and the secondary LLP decay vertex lie within the fiducial volume and are distinctly resolvable. We define the geometric acceptance $a$ as the fraction of events satisfying these conditions, independent of detector response and reconstruction efficiency.

We evaluate $a(\beta\gamma c\tau)$ with simulations of nucleon decays over a range of decay lengths $\beta\gamma c\tau$. For each parameter point at least $10^6$ events are generated. The primary nucleon decay vertex is sampled uniformly within the detector's fiducial volume  and the LLP is emitted isotropically. The probability density function for decay at length $l$ for each event is drawn from the distribution 
\begin{equation}
    P(l)  \mathrm{d}l = \frac{1}{\beta\gamma c\tau}
    \exp \left(-\frac{l}{\beta\gamma c\tau}\right) \mathrm{d}l    ,
    \label{eq:decay_length}
\end{equation}
where $\beta = v/c$ and $\gamma = 1/\sqrt{1 - \beta^2}$ are relativistic Lorentz factors for particle traveling with velocity $v$.
This follows from the standard particle decay law in the laboratory frame. 
The geometric acceptance is then defined as the fraction of events for which the LLP decays within the fiducial volume and the vertex separation satisfies $l > l_{\min}$, where $l_{\min}$ is the minimal resolvable displacement. Detector geometries for Super-Kamiokande, Hyper-Kamiokande and JUNO are implemented as described in the main text.

The geometric acceptance exhibits a characteristic dependence on the decay length $\beta\gamma c\tau$. For $\beta\gamma c\tau \ll l_{\min}$  the vertex separation falls below the spatial resolution and the acceptance is suppressed. The acceptance increases as the typical decay length becomes comparable to the detector scale, reaching a maximum when a substantial fraction of LLP decays occur within the fiducial volume while satisfying $l > l_{\min}$. For $\beta\gamma c\tau \gg l_{\max}$, where $l_{\max}$ is set by the detector size, the LLP typically exits the detector before decaying and the acceptance decreases.
 The peak acceptance is primarily determined by the detector geometry. We estimate the peak acceptance of $\sim 77\%$ and $\sim 84\%$ for Super-Kamiokande and Hyper-Kamiokande, respectively, considering our simplified treatment. For JUNO, the spherical geometry leads to a comparable maximal acceptance at the $\mathcal{O}(80\%)$ level.

\begin{figure}[t]
    \centering \includegraphics[scale=1]{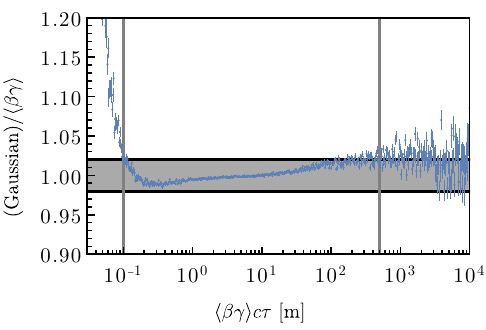}
    \centering \includegraphics[scale=1]{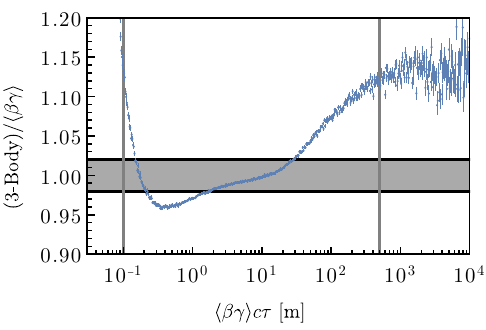}
    \caption{  
 Ratios of the geometric acceptance computed under different assumptions. (Left) The panel shows the acceptance ratio of the 2-body decay simulations including Gaussian smearing and that of fixed $\beta\gamma$ result. (Right) The panel shows the acceptance ratio of the 3-body decay simulations and the fixed $\beta\gamma$ (2-body, no smearing) result. The grey bands indicate $\pm 2\%$ variation for reference. Vertical lines at $\beta\gamma c\tau = 0.1$ and $500$~m mark the range where the acceptance exceeds $5\%$. Outside this range, statistical uncertainties that are shown here with $1\sigma$ error bars  increase due to limited event counts.   
    \label{fig:RatioAcc}}
 \end{figure}
 
We comment on the validity of the geometric acceptance approximation considering a fixed averaged value of $\beta\gamma$. For two-body decays of free protons, $\beta\gamma$ is determined kinematically by the LLP mass $m_V$. Thus, the acceptance depends only on $\beta\gamma c\tau$. In realistic detectors, however, bound nucleon decays induce a spread in $\beta\gamma$ due to nuclear effects.

For $^{16}$O, relevant to water Cherenkov detectors, nuclear effects smear the otherwise fixed two-body kinematics. Following the nuclear shell model~\cite{MayerJensen1955}, the initial bound nucleon is assigned to an $s$-shell or $p$-shell state. Binding effects are incorporated through normalized Gaussian distributions, with mean binding energies and widths $(39.0,   10.2)$~MeV for the $s$-shell and $(15.5,   3.8)$~MeV for the $p$-shell. The effective nucleon mass is then obtained by subtracting the sampled binding energy from the free nucleon mass. For the kinematics considered here, this effect is subdominant.
We also consider Fermi motion smearing, captured by proton spectral function measured in electron-carbon scattering~\cite{Nakamura:1976mb}. As an estimate, we sample $(\beta\gamma)_i$ event by event from a Gaussian distribution corresponding to a momentum width $\sigma \simeq 47$~MeV, motivated by a typical Fermi momentum $p_F\simeq 0.22$~GeV. Additionally, in-medium nuclear correlations can modify the decay kinematics through the surrounding nucleons~\cite{Yamazaki:1999gz}. We neglect these effects, since the corresponding fraction of events is expected to be subdominant.

In Fig.~\ref{fig:RatioAcc}, we compare the geometric acceptance computed with fixed $\beta\gamma$ to the result including Fermi smearing estimate. We find agreement at the level of $\lesssim 2\%$ over the range $\beta\gamma c\tau \in [0.3,   500]$~m, indicating that the fixed-$\beta\gamma$ approximation is sufficient for our sensitivity estimates.

Beyond two-body decays, multi-body nucleon decays induce a non-Gaussian $\beta\gamma$ distribution. In particular, as emphasized in Ref.~\cite{Chen:2014ifa}, matrix-element effects can further shape the kinematic distributions. To assess the impact on geometric acceptance, we simulate three-body decays by sampling $(\beta\gamma)_i$ from the full phase-space distribution. For each event, the decay length is assigned as $l_i = (\beta\gamma)_i c\tau$, drawn from the exponential law of Eq.~\eqref{eq:decay_length} evaluated at the corresponding relativistic boost. This event by event variation in $\beta\gamma$ is treated independently of the stochastic decay length sampling, and both effects are included simultaneously. As shown in Fig.~\ref{fig:RatioAcc}, the resulting three-body acceptance agrees with the fixed-$\beta\gamma$ approximation to within $\lesssim 10\%$ over the range of interest $\beta\gamma c\tau \in [0.3,    500]$~m, beyond which the acceptance ratio can exceed $5\%$. 

Our results indicate that the use of a fixed $\beta\gamma$ in the acceptance calculation is well justified at the level of our sensitivity estimates. We emphasize that the geometric acceptance $a$ is distinct from the overall signal efficiency $\epsilon$, which depends on detector response, trigger conditions, and event reconstruction. The sensitivity results presented here are therefore geometry-level estimates and should be convolved with $\epsilon$ as determined by dedicated experimental analyses.

\subsection{E. Dark Photon Parameter Space~\label{app:lim}}

We briefly discuss constraints on long-lived light
vectors and compare them with the parameter regions relevant for the
echo signatures considered in this work. It is important to emphasize
that nucleon decay experiments probe the  production  of light
vectors through baryon number-violating interactions, while the
subsequent decay of the vector is governed by its coupling to SM
states, parametrized here by the kinetic mixing parameter
$\epsilon$. This differs qualitatively from conventional dark photon
searches, in which both production and decay are controlled by the
same kinetic mixing interaction with the SM photon. Consequently,
standard dark photon searches and the displaced echo signatures
proposed here probe complementary regions of light vector parameter
space.

\begin{figure}[t]
\centering
\includegraphics{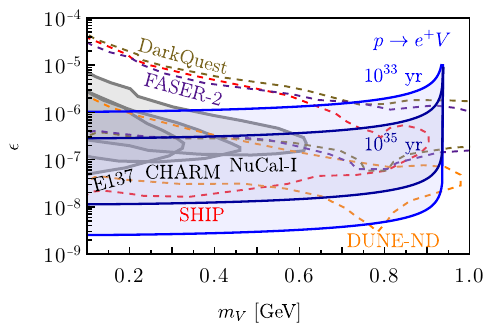}
\caption{Summary of existing constraints as well as projected sensitivity reaches for dark photons in the
kinetic mixing versus mass plane. Existing beam-dump constraints are
shaded in grey, while projected sensitivities of future experiments are
shown as coloured curves. The solid curves indicate the parameter
space accessible to Hyper-K  for  benchmark proton decay
half-life of
$T_{1/2}(p\rightarrow e^+V)=10^{33}~ \mathrm{yr}$
and
$10^{35}~\mathrm{yr}$.
See text for details. }
  \label{fig:Lim}
\end{figure}

Fig.~\ref{fig:Lim} summarizes current and projected constraints on
dark photons in the parameter region relevant for echo signals in
nucleon decay experiments. For review of dark photon  and
related light vector constraints see for example  
Ref.~\cite{Batell:2022dpx}. Particularly relevant for the present
discussion are existing beam-dump constraints from reinterpretations
of CHARM data~\cite{Gninenko:2012eq}, E137~\cite{Bjorken:1988as}, and
NuCal-I~\cite{Blumlein:2013cua}. These excluded regions are shown in
grey in Fig.~\ref{fig:Lim}. The figure also includes projected
sensitivities for several possible future experiments. The DUNE near detector (DUNE-ND)
projection is taken from Ref.~\cite{Berryman:2019dme}, while the
DarkQuest and FASER-2 sensitivities are taken from
Refs.~\cite{Berlin:2018pwi,FASER:2018eoc}, respectively. The SHIP
projection follows the recent analysis of Ref.~\cite{Zhou:2024aeu}.

Superimposed on these constraints are the parameter regions accessible
to Hyper-K for two benchmark choices of the proton decay
half-life,
$T_{1/2}(p\rightarrow e^+V)=10^{33}~ \mathrm{yr}$
and
$10^{35}~\mathrm{yr}$.
Smaller half-life  corresponds to larger experimentally accessible
regions in the $(\epsilon,m_V)$ plane. Since the projected sensitivity
of Hyper-K  reaches approximately
$T_{1/2}\simeq 2\times10^{35} ~\mathrm{yr}$ for the channels considered
here, the corresponding accessible parameter region disappears for
larger half-life. Fig.~\ref{fig:signatures} further shows that
Super-K  and JUNO probe approximately one order of magnitude smaller
half-life than Hyper-K, together with a somewhat reduced
range of
$\langle\beta\gamma\rangle c\tau$.
Consequently, the corresponding regions in the
$(\epsilon,m_V)$ plane accessible to Super-K  and JUNO are
slightly smaller than those shown for Hyper-K  in
Fig.~\ref{fig:Lim}.

Smaller values of the kinetic mixing parameter $\epsilon$ correspond
to longer vector decay lengths. As shown in Fig.~\ref{fig:Lim}, the
longest decay lengths accessible to Hyper-K extend into
regions of parameter space beyond the projected reach of various
proposed dark photon searches. On the other hand, for
$\epsilon \gtrsim 2\times10^{-8}$, future experiments such as DUNE
and SHIP are expected to probe a substantial fraction of the relevant
parameter space. An observation of an echo signature in a nucleon
decay experiment could therefore potentially be cross-checked through
complementary dark photon searches. In some regions of parameter
space, simultaneous signals in multiple experiments may be possible.
The converse implication, however, does not generally hold:
observation of a dark photon signal in experiments such as SHIP,
DUNE, FASER-2, or DarkQuest would not automatically imply an
observable echo signature in nucleon decay experiments. The
combination of displaced echo searches and conventional dark photon
searches therefore provides a means to discriminate between different
production mechanisms and underlying realizations of light vector new
physics.

\end{appendix}

\end{document}